\documentclass[12pt]{iopart}

\usepackage[italian, english]{babel}   

\usepackage[T1]{fontenc}
\usepackage[utf8]{inputenc}

\usepackage{multicol}

\usepackage{newlfont}
\usepackage{amsfonts}

\usepackage{graphicx}
\usepackage{longtable}
\usepackage{float}
\usepackage{color}
\usepackage{bm}
\usepackage{caption}
\usepackage{fancyhdr}
\usepackage[colorlinks=true]{hyperref}
\usepackage{amssymb}
\hypersetup{colorlinks,linkcolor=black, urlcolor=blue, citecolor=blue}
\graphicspath{ {./Figure/} }

\pdfoutput=1

\begin{document}
\title[PDOZ]{PDOZ: innovative personal electronic dosimeter for electron and gamma 
$H^{*}(d)$ dosimetry}
\author{Lucia Salvi$^\star$\textsuperscript{,}$^1$\textsuperscript{,}$^2$\textsuperscript{,}$^3$, Giulia Rossi$^1$\textsuperscript{,}$^2$, Giovanni Bartolini$^1$\textsuperscript{,}$^3$, Ali Behcet Alpat$^1$, Arca Bozkurt$^4$, Mustafa Dogukan Cegil$^4$, Ahmet Talha Guleryuz$^4$ }
\address{$^\star$ {Corresponding author}}
\address{$^1$ {Istituto Nazionale di Fisica Nucleare (INFN), Section of Perugia, Perugia, Italy}}
\address{$^2$ University of Perugia, Department of Physics and Geology, Perugia, Italy}
\address{$^3$ BEAMIDE srl, Via Campo di Marte $4/$O, $06124$, Perugia, Italy}
\address{$^4$ IRADETS A.S., Teknopark Istanbul, Pendik/Istanbul, Turkiye}

\ead{lucia.salvi@pg.infn.it}
\vspace{10pt}

\begin{abstract}
\newline
The personal (or active) electronic dosimeters (PEDs) are devices used to determine the individual exposure to ionizing radiations and they are employed in hospitals, research laboratories and nuclear power plants. The PDOZ project is a personal electronic dosimeter able to detect, discriminate and measure the delivered dose by beta particles and gamma rays. 
In this paper, several Monte Carlo simulations are described. The first one is regarding the ICRU sphere, \cite{ICRU39}, \cite{ICRU51} implemented to evaluate the ambient dose equivalent, $H^{*}(10)$, and the fluence-to-dose equivalent conversion coefficients for gamma rays and beta particles. The second simulation is carried out to study the prototype dosimeter response to gamma rays and beta particles and, also thanks to previous one, to obtain the conversion curve necessary to calculate the ambient dose equivalent from the silicon photomultipliers counts. In the last one, instead, the performance of a prototype dosimeter, composed by a small plastic scintillator coupled to two SiPMs, is evaluated and a simulation with different radioactive sources is made whose results are compared with the experimental measurements. All simulations are carried out by Geant4 including the optical photon transport.
\end{abstract}

\vspace{2pc}

\maketitle

\section {Introduction}
In recent years, dosimetric measurements in the various fields of application of ionizing radiation have become increasingly necessary. Nuclear applications, in fact, are not only for electrical power generation or medical applications concerning diagnostic imaging and radiotherapy or hadrontherapy treatments. Ionizing radiation is also used for sterilization of medical and pharmaceutical products and for irradiation of food to improve its preservation for long periods of time. Simultaneously with these applications, dosimetric systems have been developed to control radiation processes and determine the amount of energy released from radiation into matter \cite{Shani}. Furthermore, the radiation received by exposed people during and after catastrophic events like Fukushima, Chernobyl or in a radiological accident must be accurately measured. The precision and reliability of the measurement even in this harsh radiation environment situations are of paramount significance for decision making after such an unfavorable event and on protective action \cite{BARQUINERO2021106175}. These events have shown that most of state of the art of commercial PEDs show their limitations in particular in linearity of their response to wide energy mixed radiation fields (i.e. electrons, photons and neutrons). We design PDOZ, that will use both organic and inorganic scintillators, to be a reliable PED with linear response in mixed fields of radiation of wide energy ranges. A non - exhaustive list of applications of PDOZ can be as: in medical imaging well logging, homeland security, marine and space exploration, and high energy physics (HEP).\\
Nowadays, there are many different dosimetric methods, each of which provides information on the energy and dose absorbed by the medium on which the ionizing radiation impacted. \\
Personal dosimeter currently on the market can detect gamma rays and neutrons; gamma rays and thermal and fast neutrons; gamma rays and beta particles; gamma, X - rays and neutrons. For example the EPD - N2 dosimeter, developed by Thermo Fisher Scientific \cite{ThermoFisher}, can detect gamma radiation (and also x - rays) and thermal neutrons in the energy ranges $20$ keV - $10$ MeV and $0.025$ eV - $15$ MeV respectively with an energy response ranging from $\pm 20\%$ to $\pm 50\%$ for gammas (it depends on the energy) and $\pm 30\%$ for neutrons.     
Other example to personal electronic dosimeters with multi - detector capabilities is the NRF series, namely NRF30, NRF31 and NRF34, developed bu FujiElectric \cite{FujiElectric}. The NRF30 can detect gamma and X - ray in the energy range from $30$ keV to $6$ MeV. The NRF31 instead can detect both gamma (X) - rays and neutrons in the energy ranges $30$ keV - $6$ MeV and $0.025$ eV - $15$ MeV respectively. In the end, the NRF34 can detect gamma (X) - rays and beta particles in the energy ranges $30$ keV - $6$ MeV and $0.2$ MeV - $0.8$ MeV (mean energy). For all of them the energy response is $\leq\pm 20\%$ for gamma (X) - rays, $\leq\pm 30\%$ for beta particles and $\leq\pm 50\%$ for neutrons. None of these personal electronic dosimeters can detect gamma - rays, beta particles and neutrons at the same time.    

The PDOZ is a project consisting in a dosimeter development that, in its final version, will be able to detect and discriminate three different kinds of particles: beta, gamma and neutrons. For this purpose three scintillators, a plastic one and two distinct crystals, will be employed and two silicon photomultipliers, SiPMs, will be placed under the bottom surface of each of them. Each scintillator is specific to one type of particle to be detected: the plastic one is employed to measure the dose delivered by beta particles, whereas two distinct crystals measure the dose delivered by gamma particles and neutrons. The silicon photomultipliers are located in the bottom side of the scintillators in order to detect the light produced when a particle enters inside them and releases some energy. This energy excites the scintillator atoms to luminescence and during their return to ground state they emit light photons, in our case wavelength in blue light region, which are then collected by silicon photomultipliers. Scintillators play a crucial role as radiation detection material. They indirectly detect radiation and are usually coupled with a photo-sensor that reads the energy deposited in scintillator which are converted into light photons. Scintillators are classified into organic and inorganic, and the scintillator type used in a radiation detector is determined by the type of radiation particle to be measured as well as the purpose of radiation detection. Plastic scintillators are organic scintillators while inorganic scintillators are primarily ionic solids and composed of high - density crystals. Inorganic scintillators can then be classified into two categories, single - crystals and polycrystalline ceramics, with the former typically exhibiting better optical properties at the expense of fabrication costs \cite{Park_2017}, \cite{Cherepy}. \\
The electronic circuit is implemented to consider the light as a signal when both the SiPMs have a pulse over threshold in coincidence. When a coincidence occurs it is interpreted as a count in the device. Compared to traditional PhotoMultiplier Tube (PMT) sensors, small sized SiPMs sensors are widely used in personal dosimeters, have lower operating voltages and consume less power. PDOZ will have several dosimetry applications where the user can observe the real time dose and the dose deposited by each type of particle. It can be used in various places, for instance hospitals, research centers, nuclear power plants, laboratories and in every environment where radiations are present.\\
In this work, the study of the fluence to ambient dose equivalent curve, the dosimeter response to gamma and beta particles and the conversion curves are described in the ICRU simulation and single dosimeter response simulation. In the end, the performance study of a single dosimeter with different radioactive sources is carried out, and the results are compared and validated with experimental data.

\section {Materials}
\begin{figure} [H]
\centering
\includegraphics[width = 12 cm]{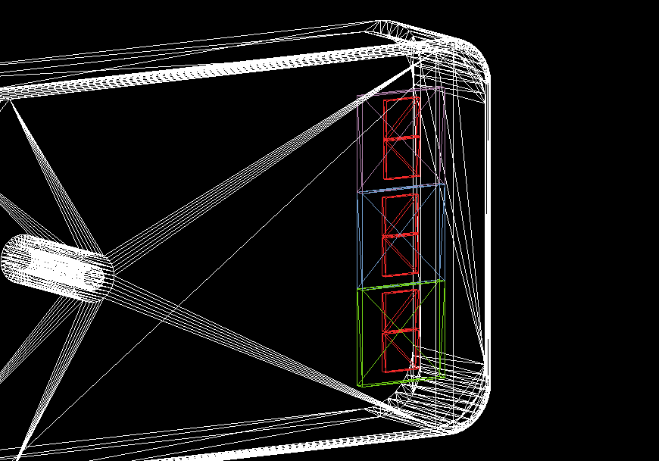}
\caption{\textit{Geometry of the PDOZ. The scintillators are in purple, blue and green, whereas the SiPMs, two for each scintillator, are in red.}}
\label{PDOZ}
\end{figure}

The PDOZ dosimeter will be able to measure the absorbed dose released from different kinds of sources. For this reason, the final version includes three separate scintillators:
\begin{itemize}
    \item plastic scintillator, the BC408  \cite{BC408}, for beta particles; 
    \item crystal, CsI(Tl) \cite{CsI}, for gamma rays;
    \item another \textsuperscript{$10$}B or \textsuperscript{$6$}LiF coated crystal, (to be decided), for neutrons.
\end{itemize}
The three scintillators are put at the same height inside the PEDs external box together with their electronic circuit for the detection. Coupled to the bottom of each scintillator there are two silicon photomultipliers, SiPM: ASD-NUV1C-P \cite{SiPM}, to detect the light produced by entering particles. Figure~\ref{PDOZ} shows the device with the three scintillators.
To study the performance of a single scintillator dosimeter (using BC408) three calibrated laboratory radioactive sources have been used to collect data:
\begin{itemize}
    \item  $^{90}$Sr having an activity A $= \left(3.7 \pm 20\% \right)$ kBq and half - life $\tau_{1/2} = 28.98$ years;
    \item $^{60}$Co with an activity A $= \left(37 \pm 5\% \right)$ kBq and half - life $\tau_{1/2} = 5.27$ years;
    \item $^{137}$Cs having an activity A $= \left(33 \pm 20\% \right)$ kBq and half - life $\tau_{1/2} = 30.08$ years.
\end{itemize}
The simulations of the ICRU sphere and those carried out to study the response of a single scintillator to gamma and beta particles and to the radioactive sources are implemented thanks to the Geant4 \cite{GEANT4} toolkit, that is used to create simulations of particles and radiation passage through the matter and any kind of setup or detector.

\section {Conversion curves calculation}

\subsection {Definitions of ICRU sphere and H*(10)}

\begin{figure} [H]
\centering
\includegraphics[width = 12 cm]{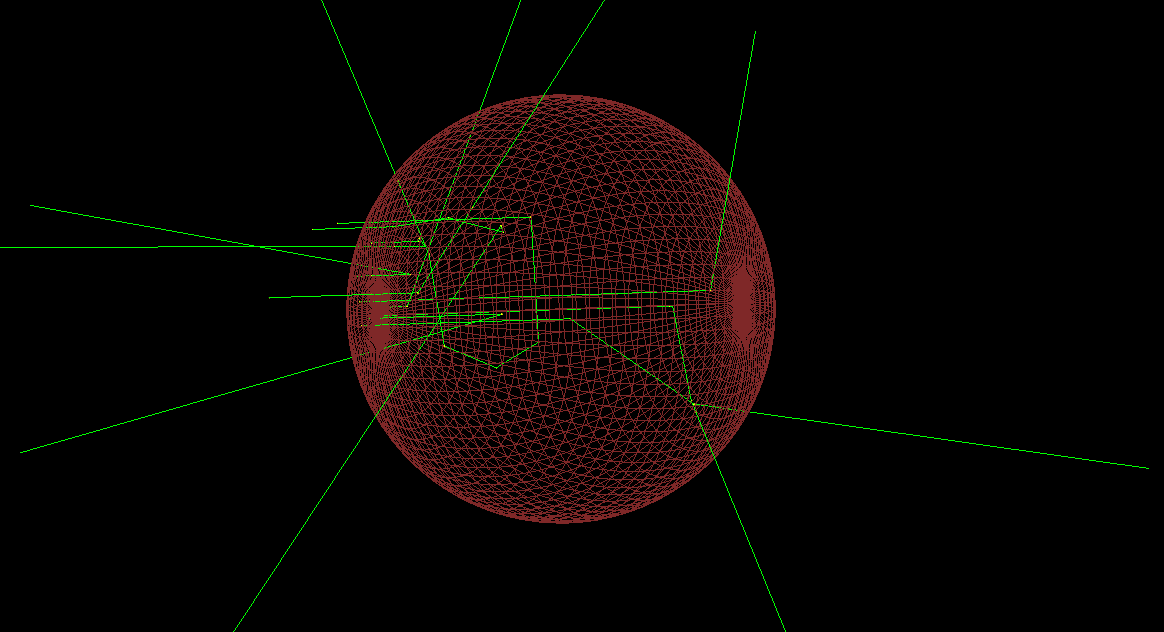}
\caption{\textit{Monte Carlo simulation of a parallel gamma rays beam at $100$ keV, impinging on a ICRU sphere where different interactions take place.}}
\label{icru}
\end{figure}

The definitions of ICRU sphere and ambient dose equivalent, $H^{*}(10)$, employed are the ones provided by the International Commission on Radiological Protection. The ICRU sphere is a sphere with unity density, $1$ g/cm$^{3}$, a diameter of $30$ cm and made of tissue-equivalent material, so $76.2\%$ oxygen, $11.1\%$ carbon, $10.1\%$ hydrogen, $2.6\%$ nitrogen \cite{ICRU39}, \cite{ICRU51} and serves to validate Monte Carlo simulations used to trace the conversion curve from count/time reading to ambient dose equivalent. Figure~\ref{icru} illustrates the ICRU sphere simulation. 
The $H^{*}(10)$, instead, is: "the ambient dose equivalent at a point in a radiation field is the dose equivalent that would be produced by the corresponding expanding and aligned field in the ICRU sphere at a depth of $10$ mm on the radius vector opposing the direction of the aligned field \cite{ICRU39}, \cite{ICRU51}". 

\subsection {ICRU Simulations}
To calculate the fluence-to-dose equivalent conversion coefficients both for gamma and beta particles, Monte Carlo simulations are implemented with Geant4. An ICRU sphere is simulated together with a sensitive volume composed by a tissue-equivalent cube having dimensions $1 \times 1 \times 1$ mm$^{3}$ and placed at a depth of $10$ mm on the radius vector opposing the direction of the aligned field for the particles. The source distribution is peaked on the axis where the cube is placed in order to reduce the simulation time without decreasing the statistic. The events are generated according to Equation~\ref{eq1}, \cite{FERRARI},

\begin{equation}
    r = R \cdot \xi^{\frac{1}{1 - \alpha}} 
    \label{eq1}
\end{equation}

in which $r$ is the radial coordinate, $R = 15$ cm represents the beam radius, $\xi \in \left(0,1\right)$ stands for a random number and $\alpha$ indicates a constant parameter set at $0.5$ as reported in the article \cite{FERRARI}.
The scored quantities must be weighted with the statistical weight $w$ defined in Equation~\ref{eq2}.

\begin{equation}
    w = \frac{2}{1 - \alpha} \cdot \frac{r^{1+\alpha}}{R^{1+\alpha}}
    \label{eq2}
\end{equation}

Several beams of $10^{7}$ monoenergetic gamma rays are generated at different energy points inside the range of $0$ keV to $2000$ keV. 

The same procedure is then applied to beta particles. Since electrons are low penetrating particles, the ambient dose equivalent, $H^{*}(d)$, is calculated at a depth of $d = 0.07$ mm as it is recommended by the ICRP standard for low penetrating radiations \cite{ICRP74}. For this reason, the geometry is changed and a box of $0.01 \times 0.01 \times 0.01$ mm$^{3}$ is placed at a depth of $0.075$ mm. In this case $10^{7}$ monoenergetic beta particles are generated at different energy points inside the range of $100$ keV to $5000$ keV.

\begin{figure} [H]
\centering
\includegraphics[width = 14 cm]{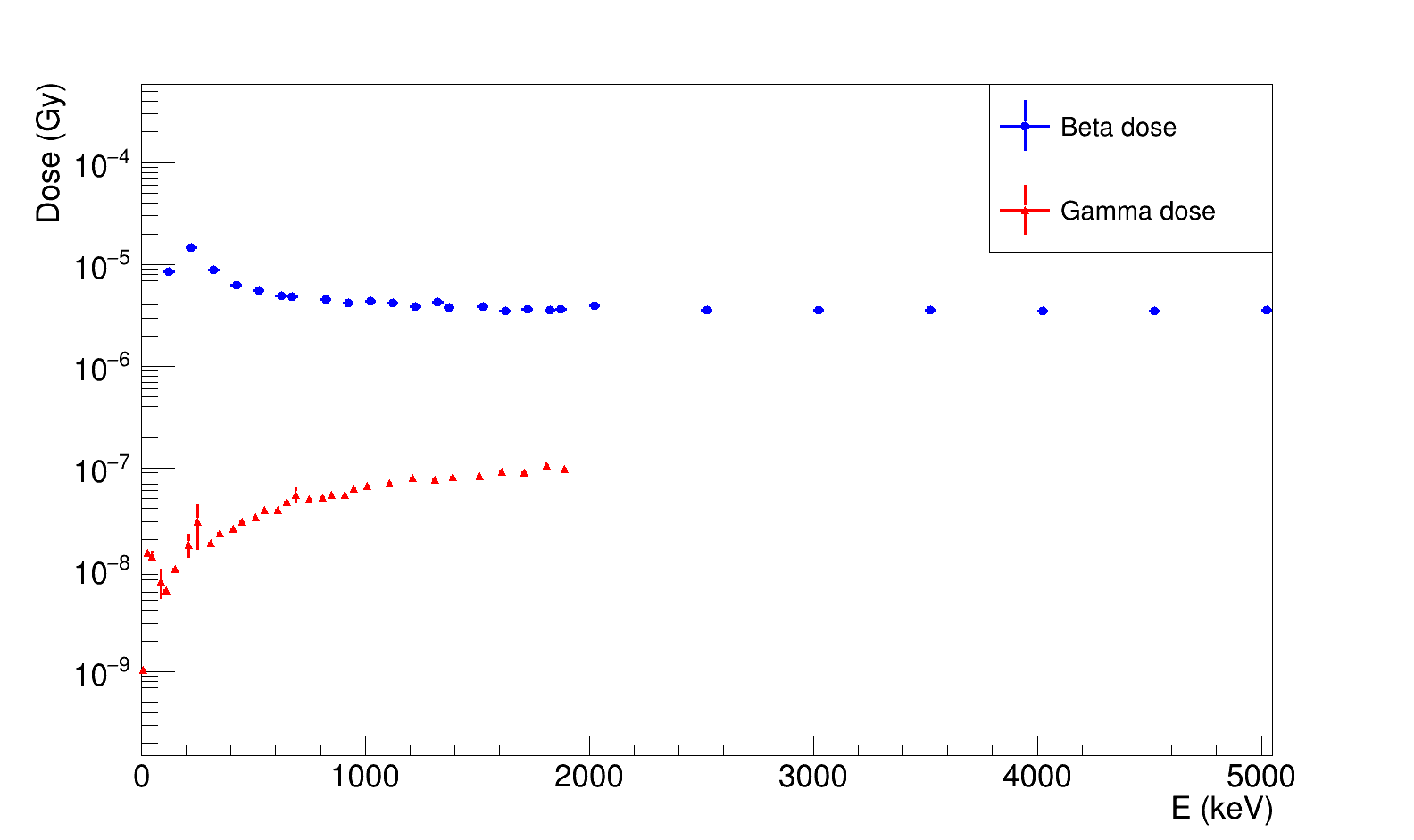}
\caption{\textit{The dose deposited in ICRU box for gamma particles in red and for beta particles in blue.}}
\label{fig1_2}
\end{figure}

In every simulation, for each energy, the absorbed dose in the ICRU box is recorded. Figure~\ref{fig1_2} shows the dose in the ICRU box volume.

\subsection {Fluence to ambient dose equivalent}
\begin{figure} [H]
\centering
\includegraphics[width = 15.5 cm]{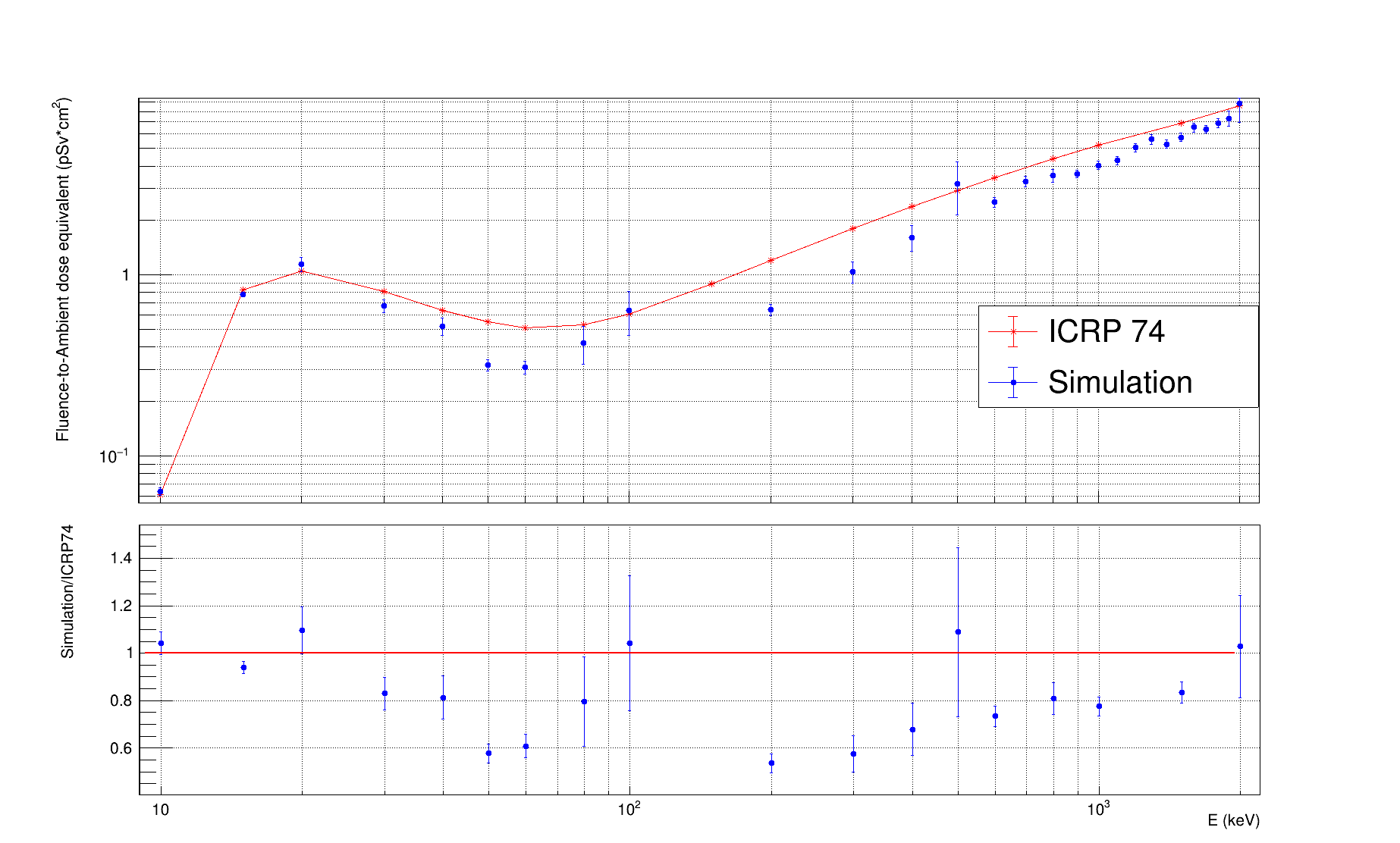}
\caption{\textit{In the first picture, the comparison between the conversion coefficients from our simulation and the ICRP74 values for gamma particles in logarithmic scale. In the second one, the ratio between the conversion coefficients from the simulation and the ICRP74 values.}}
\label{fig2_1}
\end{figure}

\begin{figure} [H]
\centering
\includegraphics[width = 15.5 cm]{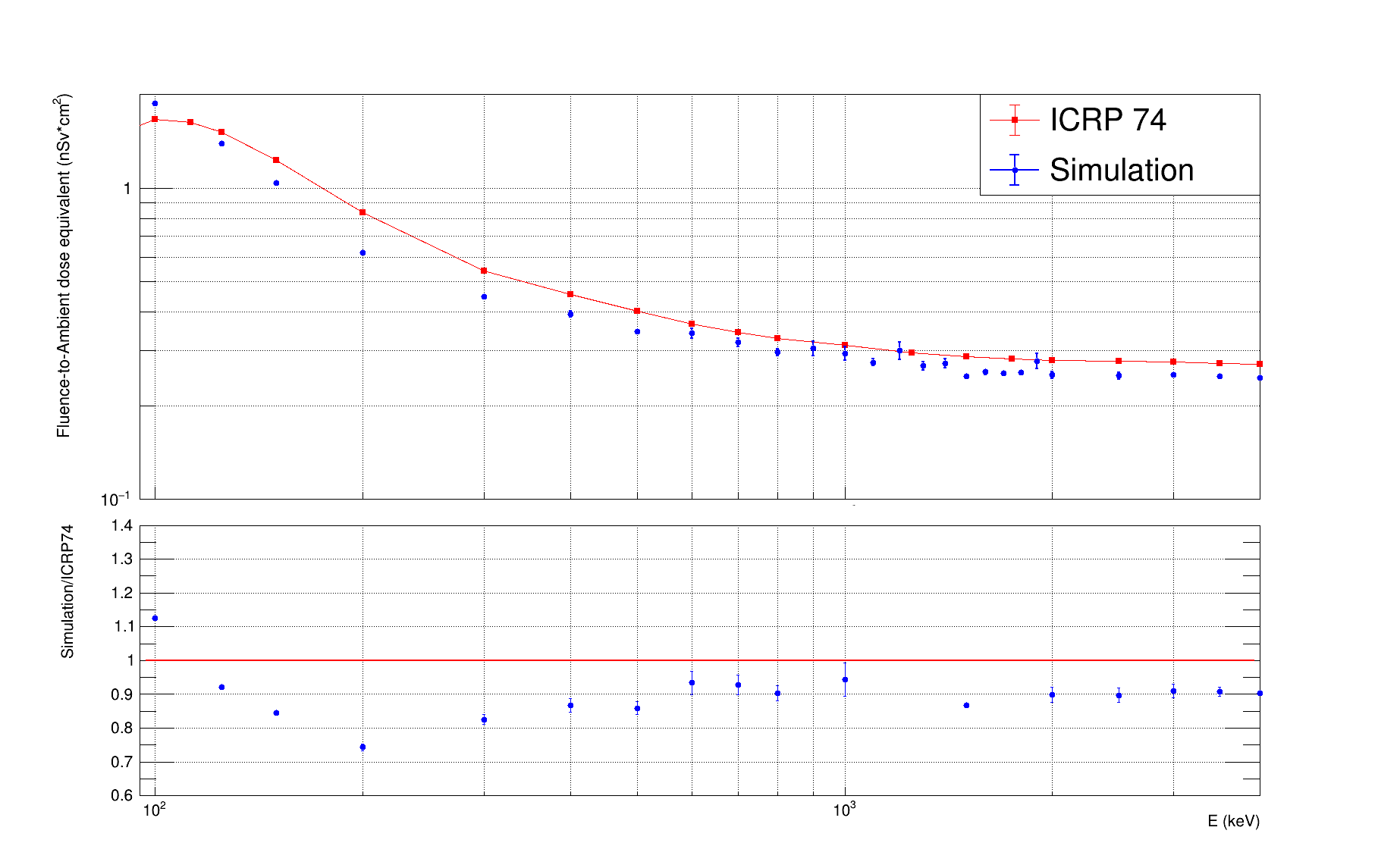}
\caption{\textit{In the first picture, the comparison between the conversion coefficients from our simulation and the ICRP74 values for beta particles in logarithmic scale. In the second one, the ratio between the conversion coefficients from the simulation and the ICRP74 values.}}
\label{fig3_1}
\end{figure}

To calculate the fluence-to-dose equivalent conversion coefficients, the fluence is determined with Equation~\ref{eq3}
\begin{equation}
    F = \frac{N}{A}
    \label{eq3}
\end{equation}
in which $F$ is the fluence measured in cm$^{-2}$, $N$ stands for the number of generated events and $A$ represents the source area which is a circle of radius $2$ cm. The source radius is now set to $2$ cm because also the dosimeter dimensions are smaller than those of the ICRU sphere. The fluence-to-dose equivalent conversion coefficients are found with Equation~\ref{eq4}
\begin{equation}
    X = \frac{D}{F}
    \label{eq4}
\end{equation}
where $X$ is the fluence-to-dose equivalent conversion coefficient measured in pSv$\cdot$cm$^{2}$ for gamma rays and in nSv$\cdot$cm$^{2}$ for beta particles, $D$ represents the dose and $F$ indicates the fluence. Figure~\ref{fig2_1} and Figure~\ref{fig3_1} report the results.



\subsection {Plastic scintillator dosimeter simulation} \label{plastic}
\begin{figure} [H]
\centering
\includegraphics[width = 7.5 cm, height= 4 cm]{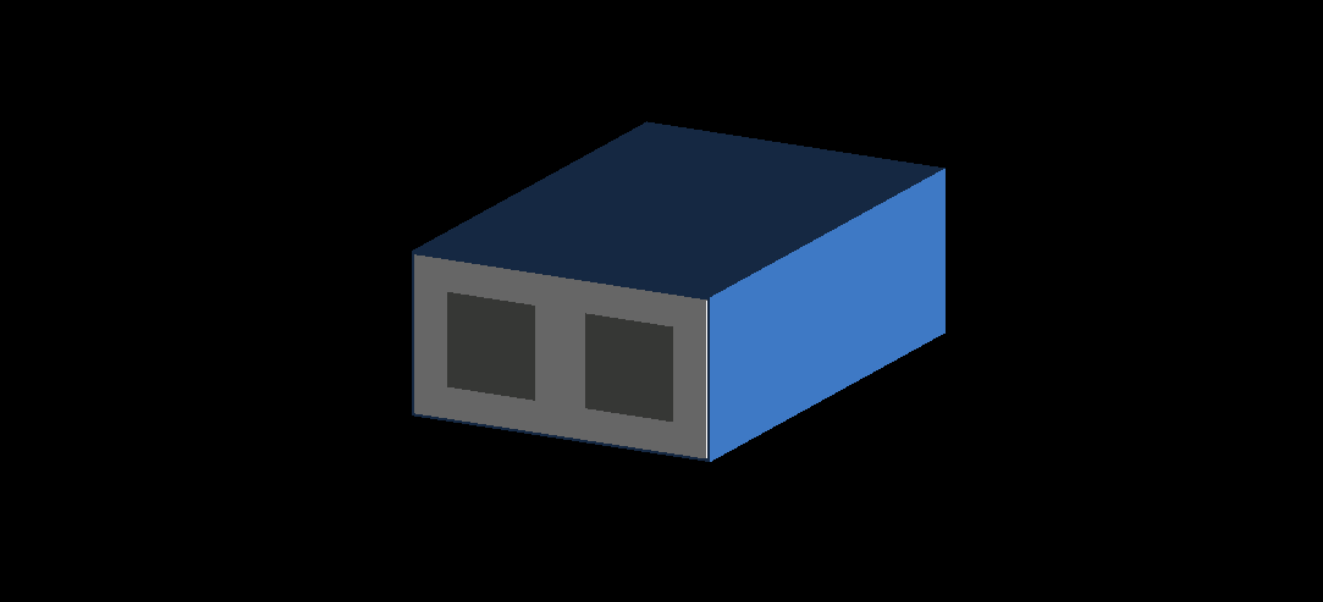}\quad\includegraphics[width = 7.5 cm, height= 4 cm]{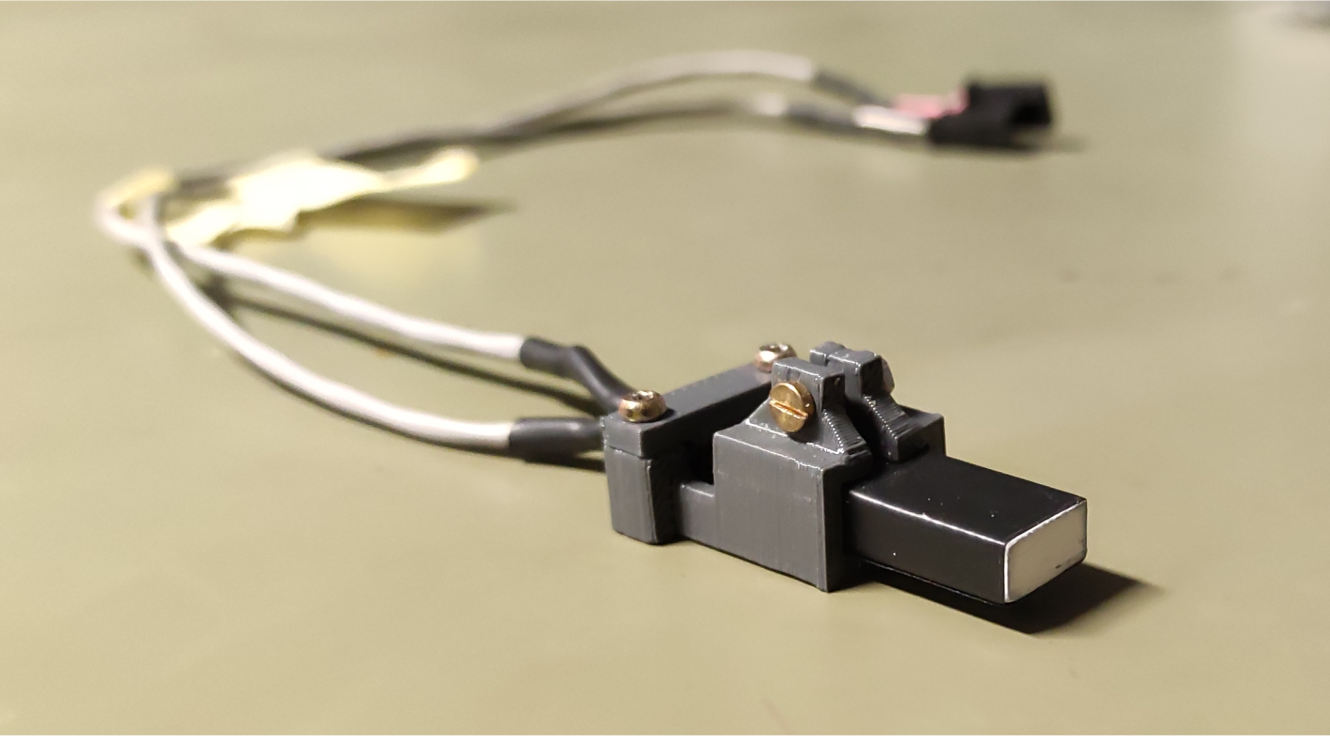}
\caption{\textit{In the left picture the Geant4 dosimeter simulation: in grey the two holes below for SiPMs and their cables, in blue the teflon foil. In the right image the experimental device with plastic scintillator.}}
\label{prototipo}
\end{figure}

In order to study the performance of a single dosimeter, the device energetic and geometrical efficiencies when subjected to different types of radiation are simulated. The incoming particle hits the scintillator, the energy deposited in it excites the scintillator molecules which then release some energy in form of optical photons. The wavelength of the optical photon released by BC408 in use is peaked around $420$ nm where the coupled SiPM's quantum efficiency is at its maximum. To evaluate the dose, the conversion factors, from the count per seconds to Sv/h, are obtained by studying the dosimeter response to different kinds of ionising radiation. The simulation geometry employed has to match that of the experimental prototype, so it involves a $10 \times 5 \times 15$ mm$^{3}$ plastic scintillator, the BC408, two holes for the silicon photomultipliers and their cables and a $100$ $\mu$m foil of teflon wrapping to avoid optical photons to escape from scintillator. In order to reproduce the behaviour as realistic as possible, the photon detection efficiency of the silicon photomultipliers is taken into account. Figure \ref{prototipo} shows the simulation of the single dosimeter geometry simulation and the experimental device.

The silicon photomultiplier, or SiPM, is a solid state photodetectors consisting of an array of several (hundreds or thousands) integrated single-photon avalanche diodes, SPADs, which are called microcells or pixel. When a photon is detected, the SPAD produce a large electric output signal because of the internal avalanche multiplication. In a SiPM it is feasible to count every hit SPAD independently and so it is possible to detect and count photons with high resolution and with single-photon sensitivity.\\ 
The SiPMs are used to detect the photons produced by a scintillator: a radiation is absorbed by the scintillator that produces photons in the near-UV or visible range. These photons are then transferred to the exit surface of the scintillator and detected by a photodetector (SiPM) in order to be converted into an electric signal. Since the SiPMs have a high photon detection efficiency (PDE), they are very suitable for this purpose \cite{Gundacker_2020}.\\
The entire process, from the production of scinintillation photons after the radiation hits the scintillator, up to the detection of these photons by the SiPM, is simulated. Also the photon detection efficiency is simulated in order to have the detector simulation as faithful as possible.\\
Three different cuts on a simulation are implemented in order to simulate three different acquisition system settings; in all of them one count takes place when in an event there is a coincidence between the signal of both SiPMs:
\begin{itemize}
    \item \textit{p.e.0}, all the coincidences (any number of photons hitting two SiPMs in coincidence) are considered, so it is required to have at least one optical photon for each SiPM; 
    \item \textit{p.e.1}, it is required to have at least two optical photons for each SiPM which means that the first photoelectron peak is eliminated from the coincidens counting (that corresponds to have only one optical photon detected by both the SiPMs);  
    \item \textit{p.e.2}, at least three optical photons for each SiPMs are required, so both the first and second photoelectron peaks are removed from the coincidence counting. 
\end{itemize}

\newpage
\subsubsection {Gamma particles simulation}
\hspace{13cm}

\begin{figure} [H]
\centering
\includegraphics[width = 14 cm]{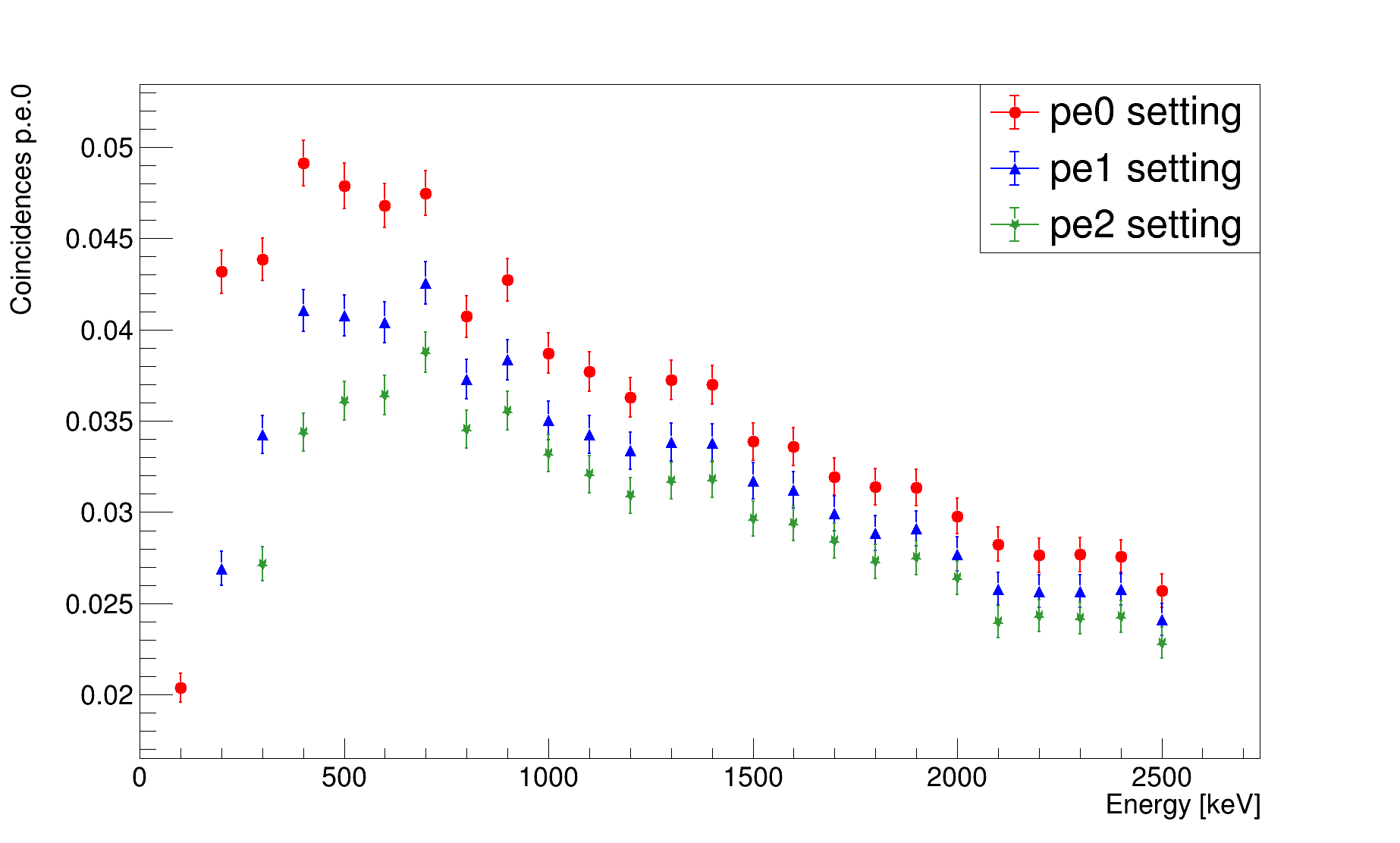}
\caption{\textit{Number of two SiPMs signal coincidences normalized to the fluence for the p.e.0, p.e.1 and p.e.2 threshold settings in red, blue and green, respectively.}}
\label{fig4_1}
\end{figure}

\hspace{-1cm} A single dosimeter, with a BC408 scintillator, is simulated with a circle-shape source having a radius, $\left(r\right)$, of $1$ cm and $10^{5}$ events are generated with the energy ranging from $100$ keV to $2500$ keV. Equation~\ref{eq5} is the fluence, in cm$^{-2}$, used to normalize the quantities:
\begin{equation}
     F = \frac{n}{\pi r^{2}} = \frac{10^{5}}{\pi} \quad\quad [cm^{-2}]
    \label{eq5}
\end{equation}
where $F$ represents the fluence, $n$ stands for the number of events and $r$ is the radius.
The dosimeter response, necessary to find the ambient dose equivalent from the SiPMs counts, is given by Equation~\ref{eq6}
\begin{equation}
    R\left(E\right) = \frac{n_{c}}{F}
    \label{eq6}
\end{equation}
in which $R\left(E\right)$ is the response, $n_{c}$ indicates the number of events with two SiPMs in coincidence and $F$ stands for the fluence. Figure~\ref{fig4_1} shows the dosimeter response in the three different settings. 

\begin{figure} [H]
\centering
\includegraphics[width = 14 cm]{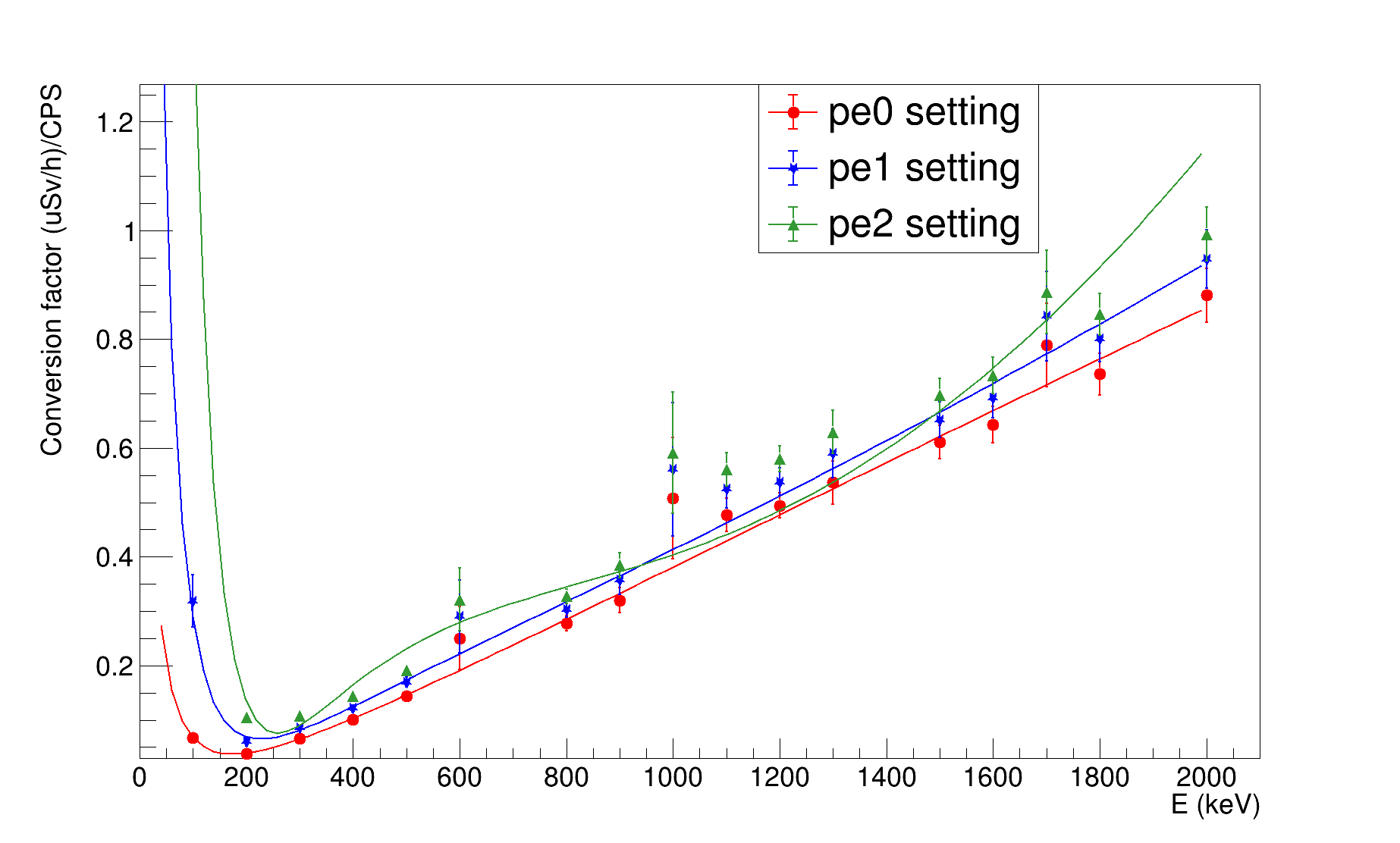}
\caption{\textit{Gamma event conversion curves for the p.e.0, p.e.1 and p.e.2 threshold settings, in red, blue and green, respectively.}}
\label{fig5_1}
\end{figure}

The conversion curve, in $\mu$Sv/CPS$\cdot$h, is estimated with Equation~\ref{eq7}
\begin{equation}
    CF = \frac{X}{R} 
    \label{eq7}
\end{equation}
where $CF$ is the conversion factor, $X$ stands for the fluence-to-dose equivalent conversion coefficient obtained from the ICRU simulation and converted in $\mu$Sv$\cdot$cm$^{2}$ and $R$ represents the response of the dosimeter. The conversion curve for the ambient dose equivalent can be estimated with Equation~\ref{eq8}, \cite{H(10)},
\begin{equation}
    f\left(E\right) = \sum_{k=1}^{K_{max}} A\left(k\right)\left(\log E\right)^{k-1}
    \label{eq8}
\end{equation}
where $A\left(k\right)$ represents a parameter, $K_{max}$ is the total number of $A\left(k\right)$ terms and $E$ stand for the energy.\\
Figure~\ref{fig5_1} illustrates the conversion curves for gamma particles fitted with Equation~\ref{eq8} in the three settings. 

\subsubsection {Beta particles simulation}
\hspace{13cm}

\begin{figure} [H]
\centering
\includegraphics[width = 14 cm]{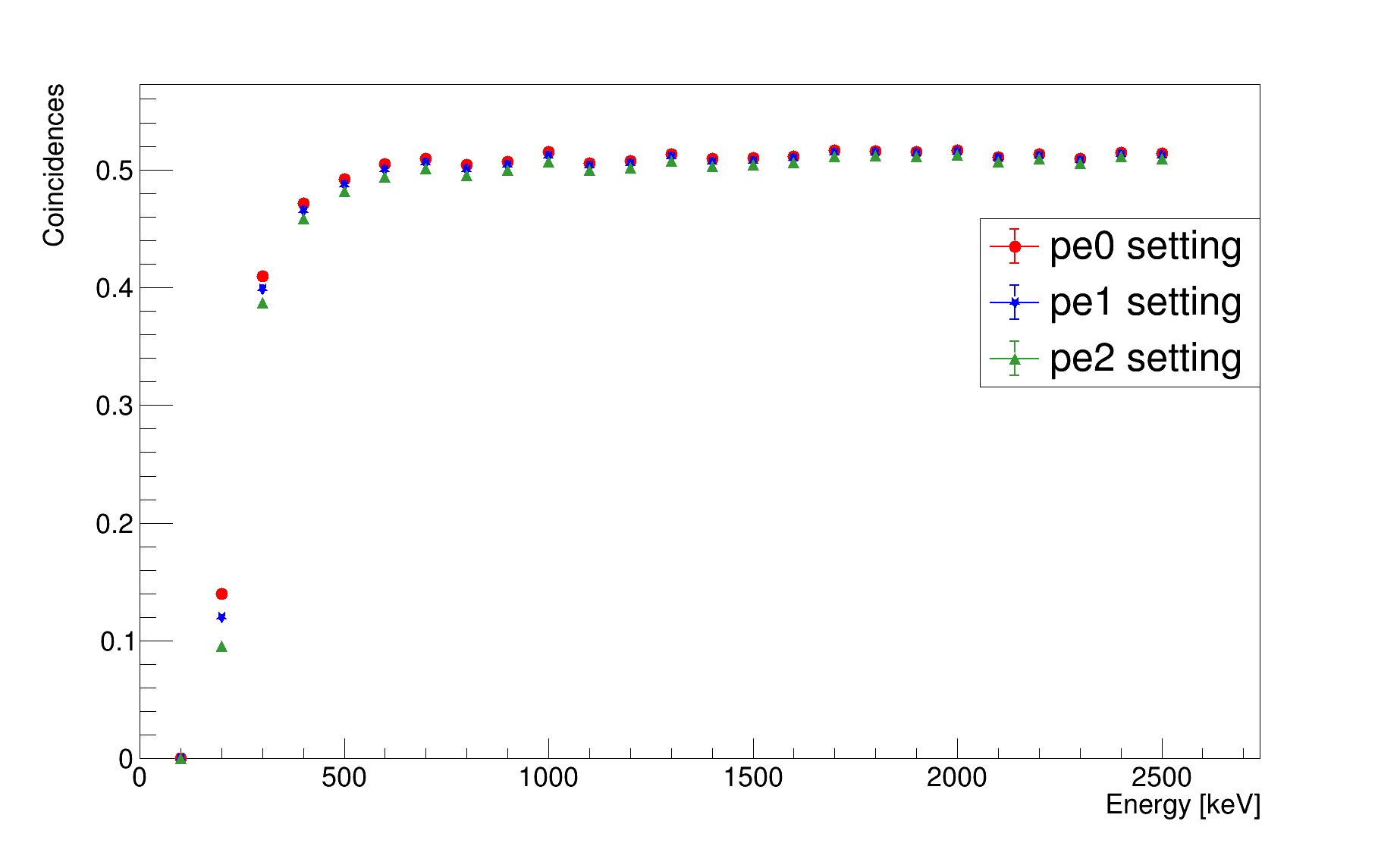}
\caption{\textit{Number of coincidences normalized to the fluence for the p.e.0, p.e.1 and p.e.2 threshold settings in red, blue and green, respectively.}}
\label{fig6_1}
\end{figure}

\begin{figure} [H]
\centering
\includegraphics[width = 14 cm]{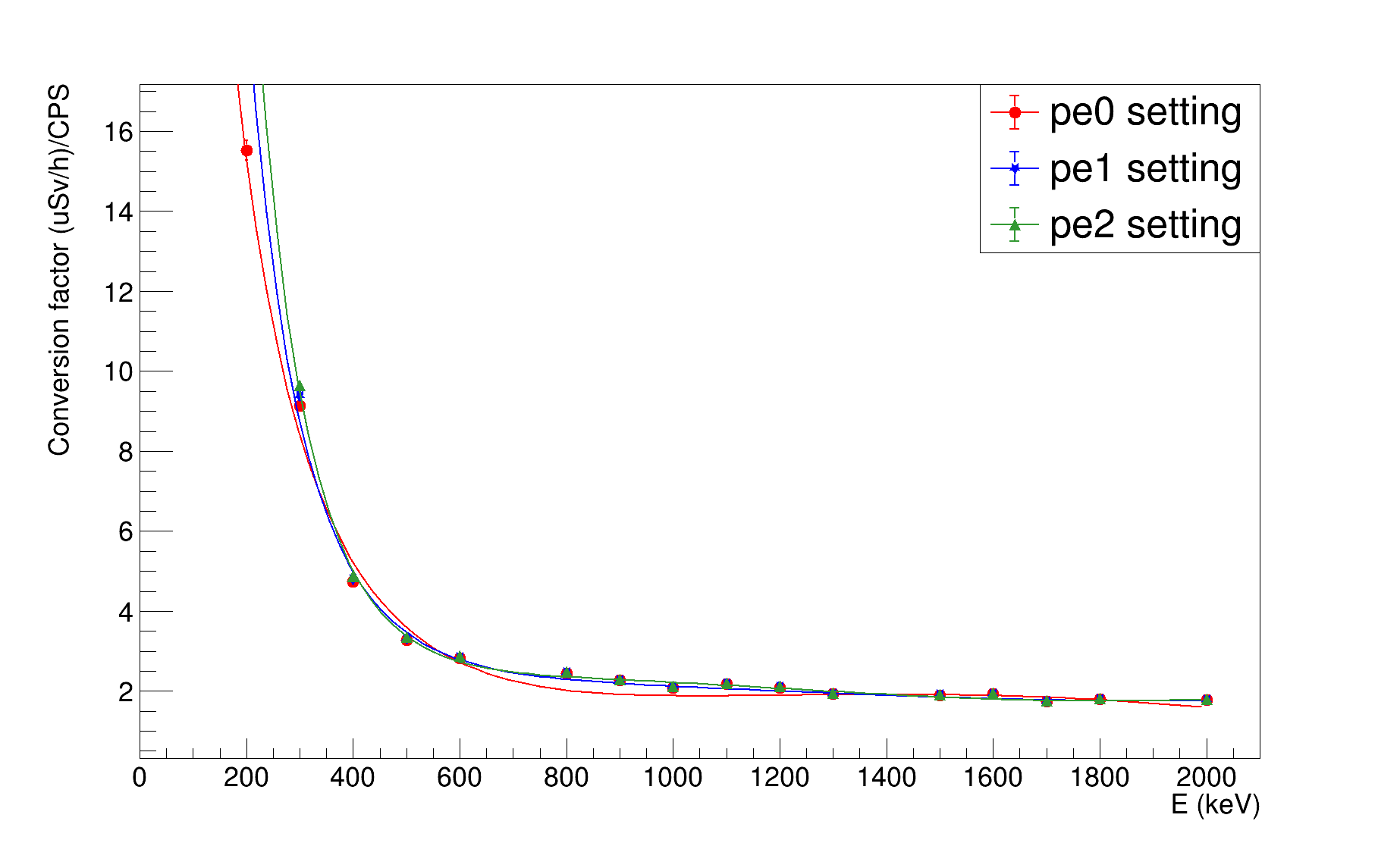}
\caption{\textit{Beta event conversion curves for the p.e.0, p.e.1 and p.e.2 threshold settings, in red, blue and green, respectively.}}
\label{fig7_1}
\end{figure}

\hspace{-1cm} The same procedure described before is applied with a beta particles source in which $10^{5}$ events are generated with energy ranging from $100$ keV to $2500$ keV. Figure~\ref{fig6_1} illustrates the dosimeter response in the three settings, whereas Figure~\ref{fig7_1} shows the conversion curve for beta particles fitted with Equation~\ref{eq8}.

\section {Results and Discussion}
\subsection {Source simulations and data analysis}
\begin{figure} [H]
\centering

\includegraphics[width = 8 cm]{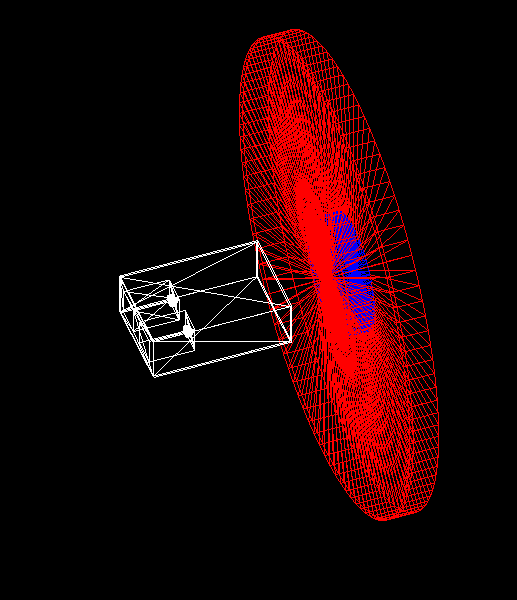}
\caption{\textit{Geometry of the plastic scintillator with two holes of depth $5$ mm each in which two SiPMs are inserted and geometry of the source. The red disk is made of plexiglass whereas the blue one of epoxy.}}
\label{fig8}
\end{figure}

In order to study the performance of one dosimeter with a single scintillator, several simulations are carried out with distinct sources whose activities are corrected taking into account the elapsed time, and, subsequently, the results are compared with the experimental data. The experimental set up involves different settings with the distance between the dosimeter and the source varying. The $^{90}$Sr, $^{60}$Co and $^{137}$Cs sources are tested in distances between $0.5 - 5$ cm.
The simulation geometry consists of a single plastic scintillator, BC404 with dimensions $10 \times 5 \times 15$ mm$^{3}$, on which a foil of teflon, $100$ $\mu$m thick, is placed. Also the source geometry is implemented in order to make comparisons with experimental data. The radioactive source is placed inside a disk of plexiglass and epoxy. The SiPMs are coupled to scintillator with EPO-TEK EJ2189 electrically conductive two-component epoxy glue \cite{epotek}. Figure~\ref{fig8} shows the experimental set up simulation. 
The sources employed are $^{90}$Sr, $^{60}$Co and $^{137}$Cs \cite{Data_Sheet} that have the following decays
\begin{itemize}
  \item[]  $^{90}$Sr $\rightarrow$ $^{90}$Y + e$^{-}$ + $\bar{\nu}$\textsubscript{e}
  \item[]  $^{60}$Co $\rightarrow$ $^{60}$Ni + e$^{-}$ + $\bar{\nu}$\textsubscript{e} + $\gamma$\textsubscript{1} + $\gamma$\textsubscript{2}
\end{itemize}
with $\gamma_{1} = 1.173$ keV and $\gamma_{2} = 1.332$ keV,
\begin{itemize}
   \item[] $^{137}$Cs $\rightarrow$ $^{137}$Ba + e$^{-}$ + $\bar{\nu}$\textsubscript{e} + $\gamma$
\end{itemize}
in which $\gamma = 661$ keV.
The sources are used to study the dosimeter detection of gamma rays and beta particles and the source spectra are those defined in Geant4.



To simulate the device behaviour, also the Photon Detection Efficiencies of the SiPMs are considered. One count is given when there is the coincidence between both SiPMs signals. The settings used to make the comparison between simulated and experimental data are the same described in section~\ref{plastic}.\\ Figure~\ref{fig10_1} shows the comparison between the simulated and the experimental data for the three different sources and settings. 

\begin{figure} [H]
\centering
\includegraphics[width = 14 cm]{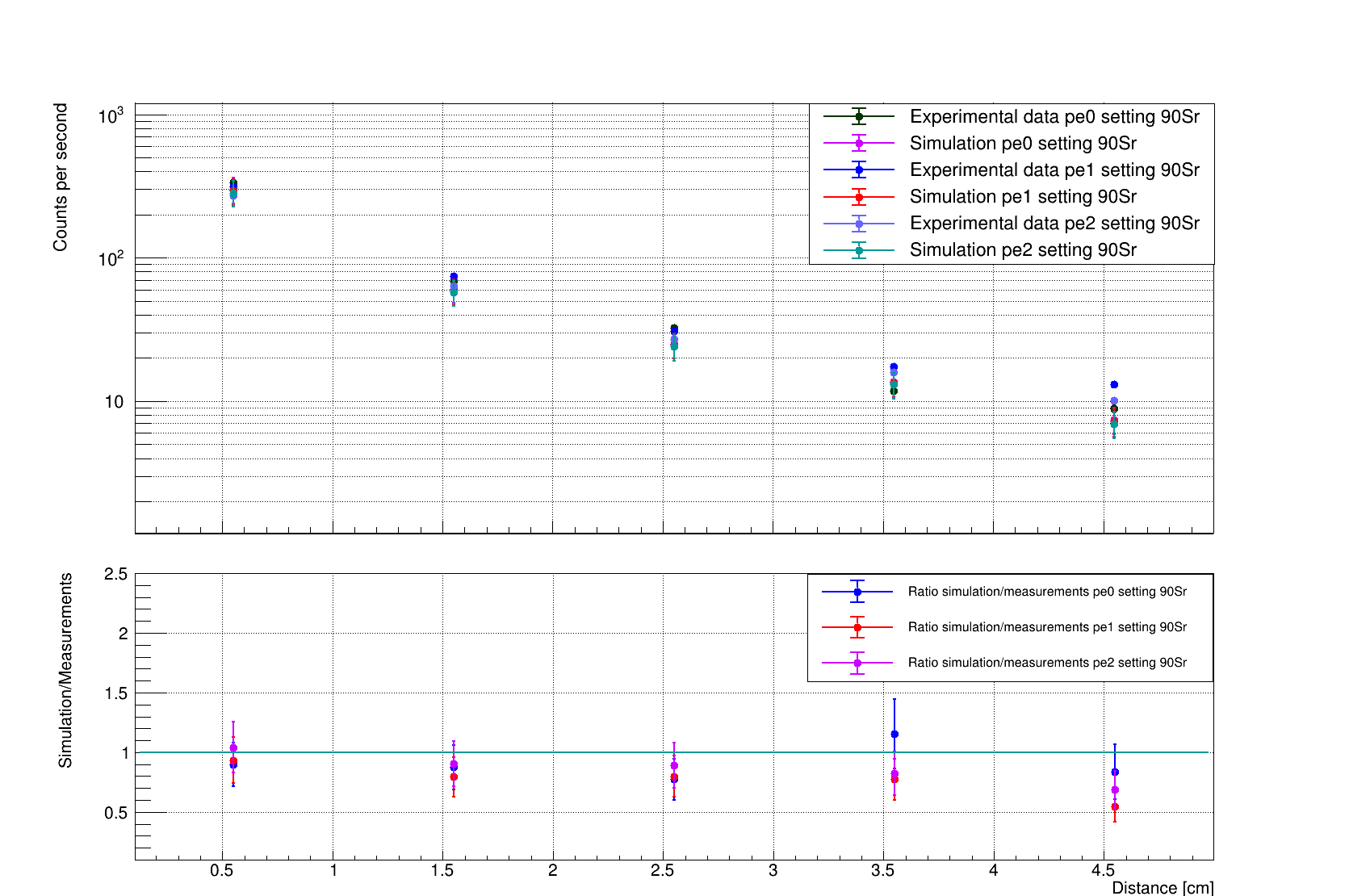}
\end{figure}
\begin{figure} [H]
\centering
\includegraphics[width = 14 cm]{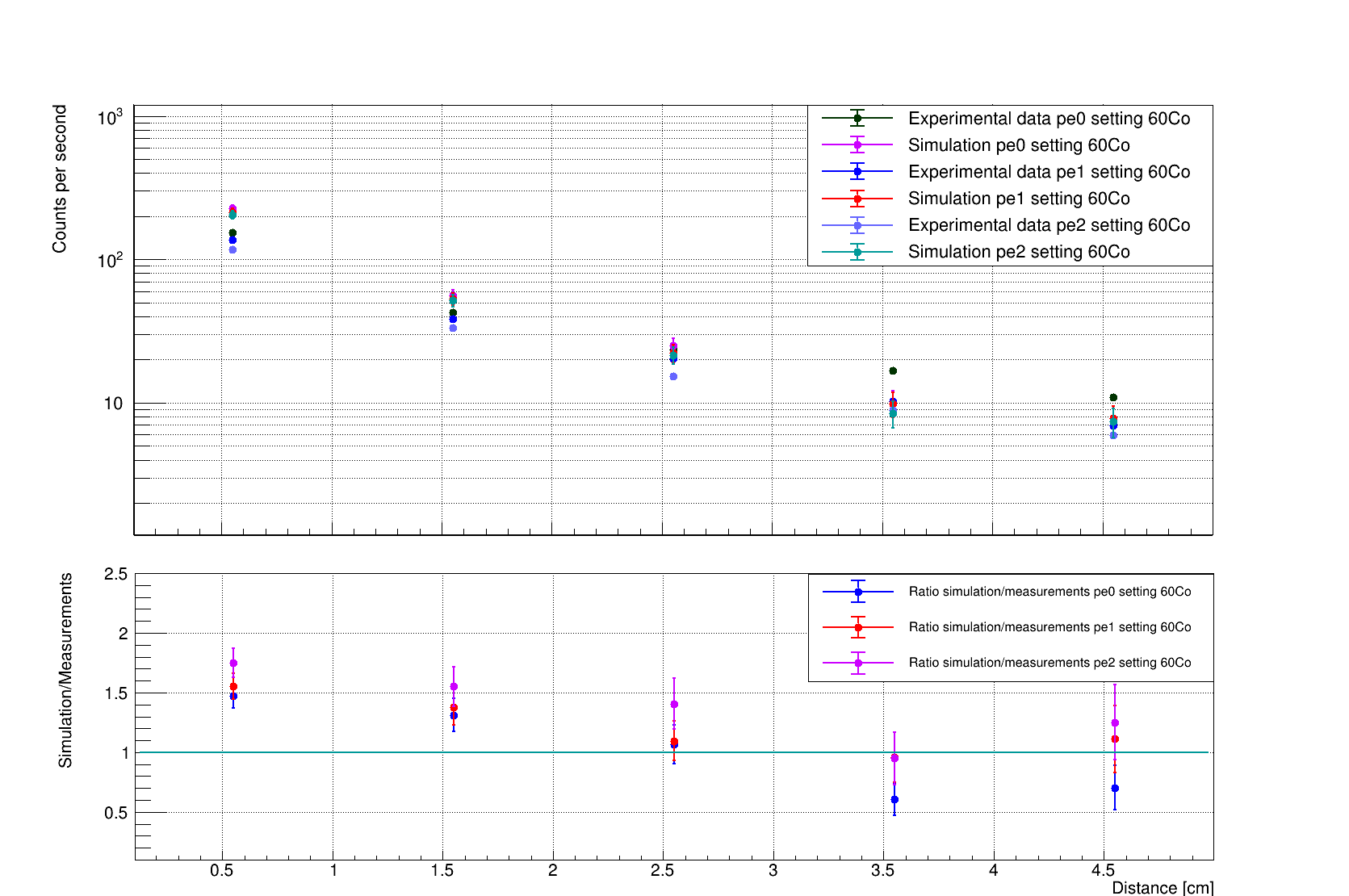}
\end{figure}
\begin{figure} [H]
\centering
\includegraphics[width = 14 cm]{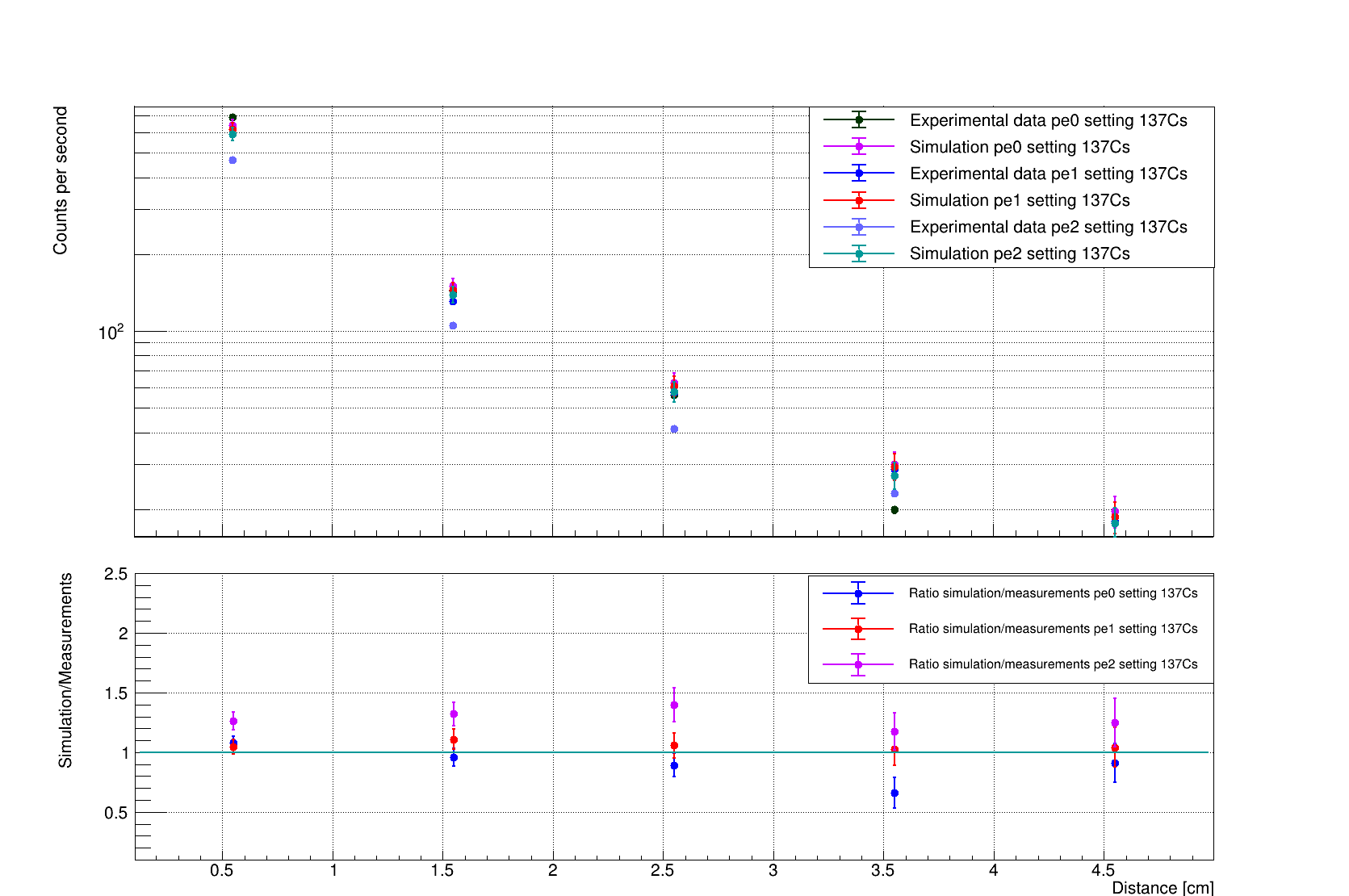}
\caption{\textit{In the first picture the comparison between simulated, in magenta, red and dark cyan, and experimental data, in dark green, blue and lilac, respectively for the p.e.0, p.e.1 and p.e.2 settings for the $^{90}$Sr source. In the second one the ratio between the simulated and the experimental data is reported for the $^{90}$Sr source.  In the other pictures the same results are shown for the $^{60}$Co and the $^{137}$Cs sources.}}
\label{fig10_1}
\end{figure}

\section {Conclusions}
In this work, the ICRU procedure suggested in \cite{CASANOVAS} has been followed to obtain the ambient dose equivalent and the fluence-to-dose equivalent conversion coefficients for gamma rays and beta particles sources. The same simulation is then implemented on the detector in order to study the performance of a single scintillator seen by two SiPMs when it is exposed to gamma ray or beta particle sources. Main goal of this study is to validate our Geant4 simulation of a single scintillator personal dosimeter (PED) following the procedure suggested by ICRU. Our PED provides coincident event counts for various configurations. The simulation creates the exact same configuration of the real experimental setup and physical processes including the optical photon transport and coincidence conditions. The agreement between the data and simulation (see counts per second versus distances curves for three different radiation sources in Figure \ref{fig10_1}) validates our Monte Carlo approach and physical parameters set. 
From these studies the conversion curves are obtained by which the dose can be evaluated. In the end, a final simulation is implemented to study the response of the device when exposed to sources of $^{90}$Sr, $^{60}$Co and $^{137}$Cs. In this last work, also experimental data are taken to do a comparison with the simulated one. The results confirm the agreement between simulated and experimental data within the error bars for both gamma rays and beta particles. Through this work, it will be possible to evaluate the dose from the counts per second using the conversion curves found.

In the end, in the new version of the PDOZ, also the response to neutrons must be studied. Therefore, in the final prototype, whose design includes particle discrimination, it will be possible to detect, discriminate and evaluate the dose released by beta particles, gamma rays and neutrons.

\section*{References}  
\bibliographystyle{unsrt}
\bibliography{refs.bib}
\end{document}